\documentclass[useAMS,usenatbib]{mn2e}

\usepackage{times}
\usepackage{amssymb}
\usepackage{lscape,graphicx}

\title[AGN environments at $z<1.5$ in the UKIDSS Ultra-Deep Survey]{AGN environments at $z<1.5$ in the UKIDSS Ultra-Deep Survey}

\author[E. J. Bradshaw et al.]{E. ~J. Bradshaw$^{1}$\thanks{E-mail:ppxeb@nottingham.ac.uk}, O. Almaini$^{1}$, W.~G. Hartley$^{1}$, R.~W. Chuter$^{1}$, 
\newauthor C. Simpson$^{2}$, C.~J. Conselice$^{1}$, J.~S. Dunlop$^{3}$, R.~J. McLure$^{3}$, M. Cirasuolo$^{3}$\\
  $^{1}$School of Physics and Astronomy, University of Nottingham, University Park, Nottingham NG7 2RD \\
  $^{2}$Astrophysics Research Institute, Liverpool John Moores University, Twelve Quays House, Egerton Wharf, Birkenhead CH41 1LD  \\
  $^{3}$Institute for Astronomy,
  University of Edinburgh, Royal Observatory, Edinburgh EH9 3HJ \\
}

\begin{document}

\date{Accepted 1988 December 15. Received 1988 December 14; in original form 1988 October 11}

\pagerange{\pageref{firstpage}--\pageref{lastpage}} \pubyear{2009}

\maketitle

\label{firstpage}

\begin{abstract}
We investigate the environments of both X-ray and radio-loud AGN within the UKIDSS Ultra-deep Survey (UDS) using deep infrared selection to sample the galaxy density field in the redshift range 1.0 $\leq z\,\leq$ 1.5. Using angular cross-correlation techniques we find that both X-ray and radio-loud AGN preferentially reside in overdense environments. We also find that both types of AGN cluster more strongly with those galaxies classified as `passive' rather than those that are actively star-forming. We infer clustering scale lengths comparable to those of passive red galaxies, suggesting that typical AGN at these epochs reside in dark-matter halos of mass $M \gtrsim 10^{13}\mathrm{M_{\sun}}$. A closer look at the small-scale environments of the AGN reveals that the neighbouring galaxies of radio-loud AGN have $U-B$ colours more skewed towards the `green-valley' and the red sequence, whereas the neighbours of X-ray AGN show no difference to the general galaxy population. This suggests that although both AGN types live in overdense environments, the radio-loud AGN may be preferentially located in more evolved cluster cores, in a similar environment to low-powered radio AGN in the local Universe.
\end{abstract}



\section{Introduction}\label{Introduction}

In the local Universe, there is a tight correlation between the mass of a galaxy's supermassive black hole (SMBH) and its stellar velocity dispersion \citep[e.g.][]{{2000ApJ...539L...9F}, {2000ApJ...539L..13G}}, which suggests a joint evolution of the two. In addition, many models of galaxy formation now invoke active galactic nuclei (AGN) feedback in order to terminate star formation in the most massive galaxies \citep[e.g.][]{{2006MNRAS.370..645B}}, although the precise mechanism is unclear. There is also increasing evidence to suggest that AGN have a significant role to play in the observed colour-bimodality of galaxies \citep[e.g.][]{{2008MNRAS.385.2049G}, {2009arXiv0909.1106S}}. The interplay between an AGN and the properties of its host galaxy may therefore provide important information to further our understanding of galaxy evolution.

Many galaxy properties are also dependent on the environment in which they reside. For example, star-formation and quasar activity have been found to depend on the local galaxy density \citep{2004MNRAS.353..713K}, whilst radio-loud galaxies are more likely to reside in the brightest galaxies in groups or clusters, rather than galaxies of the same stellar mass in different environments \citep{2007MNRAS.379..894B}. The closely related galaxy morphology-density relation was discovered by \cite{1980ApJ...236..351D}, in which, as the density of the environment increases, an increasing elliptical and S0 population is found, along with a corresponding decrease of those galaxies with a spiral morphology. It may therefore also be expected that AGN properties are dependant on the environment, and it could be the environment that triggers processes such as AGN accretion \citep{2009ApJ...695..171S}.

To gain an insight into the environments of AGN, their angular and spatial clustering have been studied extensively (although early work has been limited by small number statistics). The largest study of this kind to date is by \cite{2009ApJ...697.1634R}, who studied the autocorrelation of 30\,239 spectroscopically selected quasars in the Sloan Digital Sky Survey \citep[SDSS;][]{2000AJ....120.1579Y} with redshifts of 0.3 $\leq z\,\leq$ 2.2. A strong clustering signal was detected in the sample. There was no detection, however, of any evolution in the clustering properties of the quasars with redshift. 

It is also possible to study the autocorrelation of AGN by dividing them into categories dependent on their observational characteristics. This allows investigation into whether all AGN are derived from the same population of galaxies or whether they are fundamentally different. Research by \cite{2009A&A...494...33G} investigated the autocorrelation of X-ray sources in the redshift range 0.2 $\leq z\,\leq$ 3.0 and found a strong clustering signal at the 18$\sigma$ level. The evolution of this clustering was also studied, although there was no significant difference found between AGN below and above $z=1$. The correlation length of the AGN at $z\, \sim$ 1 was found to be similar to that of early-type galaxies, suggesting that AGN are preferentially found in massive galaxies.

There is evidence to suggest that radio-loud AGN are found in different environments to the general AGN population. Radio-loud AGN are found to be among the most clustered populations in the Universe; even more so than luminous red galaxies (LRGs). \cite{2010MNRAS.407.1078D} calculated the cross-correlation between 14,000 radio-loud AGN and 1.2 million LRGs between 0.4 $\leq z\,\leq$ 0.8 from the SDSS to find that radio-loud AGN cluster more strongly than LRGs on all scales and the luminosities of radio jets were found to correlate with the local galaxy density. \cite{2008MNRAS.391.1674W} found that a sample of radio-loud AGN cluster more strongly than a sample of radio-quiet AGN at $z \simeq 0.55$.

In general, it appears that radio AGN are associated with massive red galaxies near the centres of galaxy clusters, although this can be dependent on the power of the radio source. For example, \cite{2002MNRAS.330...17B} and \cite{2003MNRAS.343..924J} studied the galaxy cluster MS1054-03, which lies at a redshift $z=0.83$, and found an excess of radio sources within 2 arcmin of the cluster centre. Around half of the radio sources had a close pair within a projected length of $10-25\,{\rm kpc}$, which was interpreted as evidence for close or weak interactions playing a part in triggering radio activity. Similar results have also been found by other investigations, some as early as the 1980s, including \cite{1987ApJ...319...28Y} and \cite{1988MNRAS.230..131P}. \cite{1991ApJ...367....1H} investigated the relationship between the radio-luminosity dependence of environments at redshifts up to $z\, \sim$ 0.5. It appears that in the local Universe, low-power radio AGN reside in clusters, whereas high-power radio AGN reside in small groups or field environments. However, as the investigation is extended up to redshifts of $z\, \sim$ 0.5, high-power radio AGN are also found to reside in more clustered environments.

\cite{2009ApJ...696..891H} extended research on AGN clustering by comparing different wavebands in which AGN can be detected. This research differs from the previous work described above in that the cross-correlation of AGN with all other galaxies in the sample was studied, rather than looking at the AGN autocorrelation \citep[also see][]{2009ApJ...701.1484C}. AGN selected in X-ray, radio and mid-infrared wavebands within a redshift range of 0.25 $\leq z\,\leq$ 0.8 were studied. Galaxies detected in the radio band were found to be very strongly clustered (more so than those AGN detected in the other wavebands) due to the fact that they are found mainly in luminous, red galaxies (which are also strongly clustered). X-ray AGN (of luminosities $10^{42} - 10^{45}\, {\rm erg~s^{-1}}$; a similar range to those discussed in this work) were found to preferentially reside in the green valley and therefore cluster with a similar strength to those green valley galaxies. Infrared-selected AGN show the weakest clustering compared to the galaxy type in which they are found (blue and less luminous galaxies) which was interpreted by \cite{2009ApJ...696..891H} as an environmental effect triggering AGN accretion.

\cite{2009ApJ...701.1484C} divided their X-ray AGN from the AEGIS field into categories such as hardness ratio, X-ray luminosity and optical brightness and cross-correlated them with galaxies in the redshift range 0.7 $\leq z\,\leq$ 1.4. However, no difference was found in the strength of the cross-correlation between these categories. Similarly, \cite{2010ApJ...713..558K} cross-correlated a sample of $\sim$ 1550 broad-line AGN with $\sim$ 46\,000 LRGs at $z\, \sim$ 0.25 \citep[a lower redshift sample than that used by][]{2009ApJ...701.1484C}. High and low luminosity X-ray AGN were cross-correlated with the LRG sample separately, and the brighter sample was found to be more clustered. This is in contrast with what has been found at higher redshifts, which the authors suggest could be due to different mechanisms triggering AGN at different redshifts or at different halo masses. In the local Universe, \cite{2010ApJ...716L.209C} calculated the autocorrelation function of 199 hard X-ray selected AGN split into different samples to find that Type I AGN are more clustered than Type II AGN, although previously, no difference in the clustering of the two populations had been found up to redshift $z=2$ \citep[][]{{2009A&A...500..749E}, {2009A&A...494...33G}, {2009ApJ...695..171S}}.

Aside from angular and spatial correlation techniques, the environments of AGN can also be studied using simpler statistical methods such as nearest-neighbour techniques. X-ray AGN at higher redshifts 0.6 $\leq z\,\leq$ 1.4 have been found to avoid underdense environments \citep[][]{{2007ApJ...660L..15G}, {2007ApJS..168...19M}}, which is very different from what is found at lower redshifts. In the local Universe, it is generally accepted that X-ray AGN reside in a range of different environments including field galaxies, small groups of galaxies, and moderately dense environments associated with the outskirts of galaxy clusters \citep[][]{{2003MNRAS.343..924J}, {2007MNRAS.380.1467G}, {2009ApJ...695..171S}}. It appears that at lower redshifts, AGN reside in star-forming galaxies irrespective of environment \citep{2009ApJ...695..171S}.

The aim of this work is to study AGN in the X-ray and radio wavebands and their surrounding galaxies at a redshift range 1.0 $\leq z\,\leq$ 1.5 in order to determine the environments of AGN during an important epoch for the establishment of the galaxy red-sequence \citep[e.g.][]{{2007MNRAS.380..585C}, {2007A&A...476..137A}, {2010ApJ...709..644I}}. A short investigation is also undertaken into the environments of AGN between a redshift 0.5 $\leq z\,\leq$ 1.0 to see whether the correlations in this paper agree with that of previous work. In this work we cross-correlate galaxies from the UKIDSS UDS (Ultra Deep Survey) with a sample of X-ray detected AGN from the Subaru/XMM-Newton Deep Survey and also radio-loud AGN from deep radio imaging of the same field. The infrared depth and area of the UDS allows us to extend studies of AGN environments to $z>1$ without the strong biases inherent in optical galaxy selection.

Where relevant, we adopt a concordance cosmology in our analysis; $\Omega_M=0.3$, $\Omega_{\Lambda}=0.7$, $h={\rm H}_0/100 ~{\rm kms}^{-1}{\rm Mpc}^{-1} =0.7$ and $\sigma_8=0.9$. Throughout the paper, the term `XAGN' is used to refer to an X-ray AGN, and `RAGN' is used to mean a radio-loud AGN. Any distances quoted are physical distances unless otherwise stated. Section~\ref{Data} describes the data used in the cross-correlation and in Section~\ref{analysis} we present the results. Section~\ref{science} looks at the small-scale environments of AGN and finally Section~\ref{discussion} discusses the interpretation of these results and concludes the paper.


\section[]{Sample Selection}\label{Data}

\begin{figure}
\begin{center}
\hspace*{-20pt}
\includegraphics[width=8.0cm]{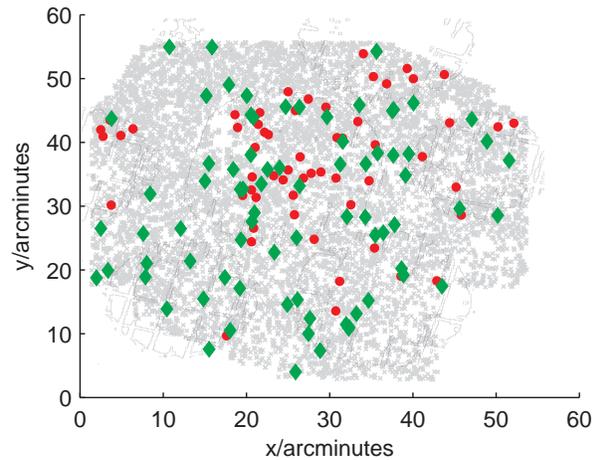} 
\end{center}
\caption{\small{Illustrating the distribution of the UDS galaxies (grey crosses) with photometric redshifts in the range 1.0 $\leq z\,\leq$ 1.5. X-ray and radio detected AGN within the same redshift range are overlaid (red circles and green diamonds respectively). The contours are drawn from the X-ray sensitivity map, representing the adopted flux limit. X-ray AGN are only expected within these contours, whilst radio AGN can be detected across the entire field. The shape of the underlying galaxy field is due to the overlapping regions of the infrared, optical and Spitzer IRAC data.}}
\label{fig:UDS}
\end{figure}

In this section we introduce the AGN and galaxy samples that were used to perform the cross-correlation analysis. We restrict our primary analysis to the redshift range 1.0 $\leq z\,\leq$ 1.5, noting that this corresponds to an important epoch in galaxy evolution: the rapid build-up of the galaxy red-sequence \citep[e.g.][]{2007MNRAS.380..585C}. The distribution of galaxies and AGN are shown in Figure~\ref{fig:UDS}.

\subsection[]{The UDS Galaxy Sample}\label{uds_section}

The UDS (Ultra Deep Survey) is one of the five surveys which make up the UKIRT Infrared Deep Sky Survey \citep[UKIDSS;][]{2007MNRAS.379.1599L}. The UDS covers an area of 0.77 square degrees, centred on the Subaru-XMM Deep Survey (SXDS), and is the deepest near-infrared survey over such a large area to date. In this work we use data from the DR3 release, reaching depths of $K_{AB} = 23.7, H_{AB} = 23.5, J_{AB} = 23.7$ (AB, $5\sigma$, $2\arcsec$). In addition to near infrared data, the field is covered by extremely deep optical data in the B, V, R, i$^{\prime}$ and z$^{\prime}$-bands from the Subaru-XMM Deep Survey, achieving depths of $B_{AB} = 28.4, ~V_{AB} = 27.8, ~R_{AB} = 27.7, ~i_{AB}^{\prime} =27.7$ and $ z_{AB}^{\prime} = 26.7$ \citep[$3\sigma$, $2\arcsec$]{{2008ApJS..176....1F}}. There is also {\it Spitzer} data reaching $5\sigma$ depths of 24.2 and 24.0 (AB) at $3.6\mu$m and $4.5\mu$m respectively from a recent {\it Spitzer} Legacy Program (SpUDS, PI:Dunlop) and U-band data taken with CFHT (Canada-France-Hawaii Telescope) Megacam ($U_{AB} = 25.5$; Foucaud et al. in prep). As the UDS data is so deep, this means that we are able to probe typical sub-L$*$ galaxies in an un-biased manner to $z\sim 3$ \citep{2010MNRAS.401.1166C}.

The galaxy sample used in this analysis is primarily based on selection in the K-band, upon which we impose a cut at $K_{AB} = 23$ to minimise photometric errors and spurious sources. We are also able to ensure a high level of completeness ($\sim 100\,$ per cent) and reliable photometric redshifts \citep[see][]{{2010MNRAS.401.1166C},{2010MNRAS.tmp.1089H}}. Bright stars are trivially removed from the catalogue by excluding objects on a stellar locus defined by 2-arcsec and 3-arcsec apertures, which is effective to $K_{AB} < 20.0$. The fainter stars are removed using a $BzK$ diagram \citep{2004ApJ...617..746D} and the criterion $(z'-K) < 0.3(B-z')-0.5$. These cuts and careful masking of bright saturated stars and surrounding contaminated regions leaves 36\,962 galaxies in our sample.

\subsubsection{Photometric Redshifts}

The photometric redshift for each galaxy in our UDS/SXDS sample has been computed by fitting the observed photometry ($UBVRi'z'JHK$ and $\rm 3.6 \mu m$ and  $\rm 4.5 \mu m$) with both synthetic and empirical galaxy templates. The fitting procedure is undertaken with a code which is largely based on the public package {\sc HYPERZ} \citep{2000A&A...363..476B}. The magnitudes obtained from the images were used to determine photometric redshifts by a $\chi^2$ minimisation over a large suite of model templates, constructed using evolutionary synthesis models of \cite{2003MNRAS.344.1000B} with a Salpeter initial mass function, and including a treatment of dust content (see \citealt{2007MNRAS.380..585C,2010MNRAS.401.1166C} for a full account). In the photometric redshift range 1.0 $\leq z\,\leq$ 1.5 we find 9\,896 K-selected galaxies in total, the spatial distribution of which are shown in Figure 1.

\subsection{X-ray Data}

The Subaru/XMM-Newton Deep Survey \citep{2008ApJS..179..124U} covers an area of 1.14 deg$^2$ and is centred at R.A. = 02h18m and Dec. = -05d. The SXDS field was mapped using XMM-Newton using seven different pointings in the 0.2-10keV band. The central pointing has the deepest exposure time (100ks), and this is surrounded by the six other pointings which have shallower (50ks) exposures. Within this survey, there were found to be 1245 X-ray point sources, of which 104 are within the redshift range 1.0 $\leq z\,\leq$ 1.5 (see below).

In order to compensate for these variations in exposure across the SXDS, we used the X-ray flux-limit map for the field in the $2-10\,{\rm keV}$ band to maximise the number of AGN used in the sample whilst keeping the flux level the same across the entire field. The flux limit was taken as $7.98\times10^{-15} \mathrm{erg~cm^{-2}~s^{-1}}$ which reduces the number of usable AGN to 62 within our required redshift range. Of these 62 AGN, 37 have spectroscopic redshifts (see Simpson et al. in preparation and Akiyama et al. in preparation) and 26 do not. See Table 1 for an overview of the various AGN samples used in this work. A K-band matching between the UDS catalogue and the SXDS was carried out in order to obtain photometric redshifts for those AGN that did not have spectroscopic data. The photometric redshift was taken from the UDS galaxy which gave the closest match between itself and the X-ray source within a 5 arcsecond radius \citep[see][]{{2003A&A...409...79F},{2010MNRAS.401..294S}}. There is an error associated with matching an X-ray catalogue to an optical or near-infrared catalogue in this fashion, as the incorrect galaxy may be assigned as the X-ray source. A rough estimate of this error can be calculated by considering the average spatial density of galaxies within the UDS and comparing this with the area covered between an X-ray source and its nearest galaxy. This calculation shows that the error on the K-band matching is 5$\%$ which corresponds to less than one of our seventeen AGN (of those with a photometric redshift only) being placed in the incorrect redshift bin.

In the redshift range 1.0 $\leq z\,\leq$ 1.5, the implied $2-10\,{\rm keV}$ X-ray luminosities are in the range $7 \times 10^{43} - 3 \times 10^{45}\, {\rm erg~s^{-1}}$ with a mean luminosity of $1.9 \times 10^{44}\, {\rm erg~s^{-1}}$. We are therefore probing `typical' X-ray AGN, around the knee in the luminosity function \citep[][]{{2009A&A...500..749E},{2000A&A...353...25M}}.

\subsection{Radio Data}

The radio data was taken from the 100-$\mu$Jy catalogue \citep{2006MNRAS.372..741S}. The SXDF was observed with the VLA using 14 overlapping pointings, in which 505 radio sources were detected over the $0.8\,{\rm deg^{2}}$ to a limit of 100$\mu$Jy at 1.4GHz. Each source is contained in the 100-$\mu$Jy catalogue along with a corresponding RA and Dec in the radio and optical wavebands. In order to create our AGN catalogue, the optical positions of the radio sources were matched to the K-band positions of galaxies in the UDS within a radius of 2 arcseconds, with the closest match chosen as the best match. 421 radio sources were matched to the UDS catalogue. Of the 421 radio sources, 72 are within a redshift of 1.0 $\leq z\,\leq$ 1.5. Of these, 11 sources have spectroscopic redshifts, and the remaining 61 have photometric redshifts. Using the definition of \cite{2008ApJ...686.1503B}, the normalised median absolute deviation ($\sigma_{NMAD}$) of redshifts is $\sigma_{NMAD} = 0.035$ with a median $\Delta z$ of -0.024; where $\Delta z = z_{phot}-z_{spec}$. From \cite{2008ApJ...686.1503B}, $\sigma_{NMAD}$ is equal to the standard deviation for a Gaussian distribution. For the 39 objects with a spectroscopic redshift $z > 1$, $\sigma_{NMAD} = 0.055$ and there is $\Delta z$ of -0.047. 

A flux density of 100$\mu$Jy at the redshift $z=1$ gives a 1.4GHz luminosity of $6 \times 10^{23}\, {\rm WHz^{-1}}$ (see Section 3.5 for a full description). This is where the star-forming galaxy radio luminosity function \citep{2007MNRAS.375..931M} starts to decline strongly. It can therefore be assumed that all radio sources above this flux limit in the catalogue which are at the redshift $z>1$ are AGN (although there may be a small amount of contamination from star-forming galaxies).

\subsection{Spatial Structure in the Spectroscopic Sample}

Potential biases may arise in our clustering analysis if the galaxies for which spectra have been obtained have spatial structure due to observational constraints. For the spectra used in this work, those objects with $R<21$ have spectra from 2dF/AAOmega, which covers the full UDS field with no gaps and minimal structure due to multiple fibre configurations. For the fainter objects with $R>21$ the spectroscopic redshifts were obtained using the Visible Multi-Object Spectrograph (VIMOS) instrument on UT3/Melipal. The programme was designed to provide the maximum number of spectra possible, using 27 overlapping masks to cover a square degree. For a more detailed description of the VIMOS mask preparation, see (Simpson et al., in preparation). Since the sky density of radio and X-ray targets with $R>21$ is relatively low ($<0.6$ per $\rm arcmin^2$) we anticipate minimal biasing due to slit placement.

In conclusion, there is no significant spatial structure in the spectroscopic sample that may bias the results of the AGN/galaxy cross-correlation. Furthermore, the primary findings of this paper are tested in Section 3.4 by comparing the results obtained from the
spectroscopic and photometric AGN samples.

\section{Probing AGN Environments with Cross-Correlations}\label{analysis}

The environments of AGN are investigated using an angular two-point cross-correlation analysis, which probes large-scale structure as well as the small-scale environment of a population of objects. The correlation function is defined as the probability of finding a galaxy in a volume element at a given separation from an object compared to a random distribution of galaxies. We calculate this using the estimator of \cite{1993ApJ...412...64L}, modified for the cross-correlation function between sample 1 and sample 2,

\begin{equation}
w(\theta) = \frac{N_{D_1D_2} - N_{D_1R_2} - N_{R_1D_2} + N_{R_1R_2}}{N_{R_1R_2}}
\end{equation}
where $N_{D_1D_2}$, $N_{D_1R_2}$, $N_{R_1D_2}$ and $N_{R_1R_2}$ are the normalised number counts of data-data pairs, data-random pairs (for both sample 1 and sample 2) and random-random pairs respectively at angular separation $\theta$. The random galaxy catalogue in this case is 50 times larger than the number of galaxies between 1.0 $\leq z\,\leq$ 1.5 in the UDS, therefore ensuring that any error introduced by the uncertainty in the data-random pairs is kept to a minimum. This estimator in equation 1 can introduce small errors to the correlation function at large scales due to the finite size of the field. This can be corrected using an integral constraint \citep[e.g.][]{1999MNRAS.307..703R} which takes the form

\begin{equation}
C=\frac{\Sigma N_{RR}(\theta)\theta^{-0.8}}{\Sigma N_{RR}(\theta)},
\end{equation}
where the sums extend to the largest separations within the field.

To allow a meaningful comparison of cross-correlation functions, we require the redshift distribution, $n(z)$, to be similar amongst the three populations (X-ray AGN, radio-loud AGN and the general galaxy population). This assumption holds with our data, as although the galaxy population does drop off slightly at high redshift, the distribution remains relatively flat over all three populations.

The AGN-galaxy cross-correlations can then be compared with the various galaxy-galaxy autocorrelation functions to provide a baseline for comparing the environments of AGN with those of normal galaxies. We refer the reader to \cite{2010MNRAS.tmp.1089H}, in which there are derived halo masses for both passive and star-forming galaxies in redshift bins relevant to this paper.

\begin{table}
\centering
\begin{tabular}{l l c c}
\hline\hline
Sample & Sub-sample & Number & No. with spec-z \\ [0.5ex]
\hline
Radio AGN & Total & 72 & 11 \\ [1ex]
\hline
X-ray AGN & Total & 62 & 37 \\ [1ex]
 & Hard X-ray AGN & 30 & 16 \\
 & Soft X-ray AGN & 32 & 21 \\ [1ex]
 & Quasar-like AGN & 38 & 19 \\
 & Seyfert-like AGN & 24 & 18 \\ [1ex]
\hline
Galaxy Sample & Total & 9\,896 & - \\
 & Passive galaxies & 1\,961 & - \\
 & Star-forming galaxies & 6\,676 & - \\
\hline\hline
\end{tabular}
\label{table:table1}
\caption{A table of the AGN and galaxy samples in the redshift range 1.0 $\leq z\,\leq$ 1.5, along with the number of galaxies in each sample and the number that have spectroscopic redshifts.}
\end{table}

\subsection{X-ray AGN correlations}\label{Xray_AGN_correlations}

\begin{figure*}
\begin{center}
\hspace*{-20pt}
\includegraphics[width=15.0cm]{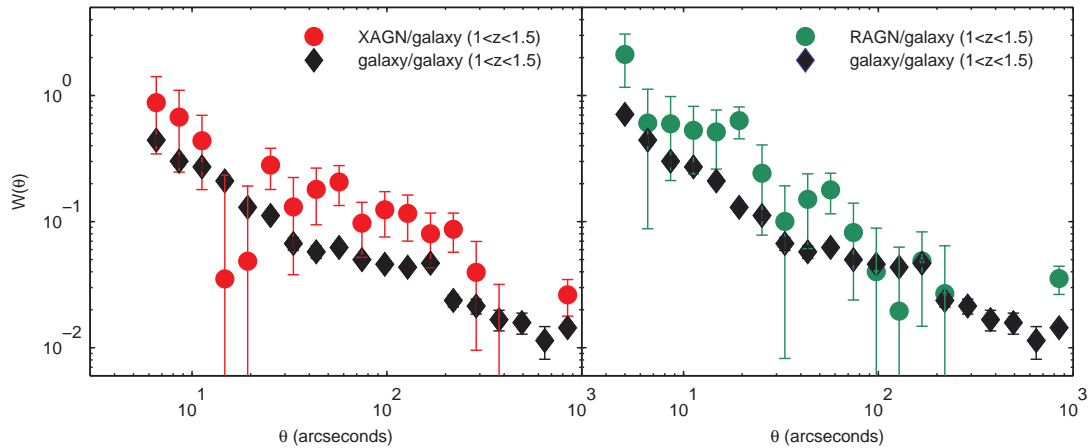} 
\end{center}
\caption{\small{The angular cross-correlation function for X-ray selected AGN between a redshift of 1.0 $\leq z\,\leq$ 1.5 compared to the galaxy-galaxy auto-correlation function within the same redshift (left) and for radio-loud AGN (right).}}
\label{fig:AGN_all_all_400000_4as_log.eps}
\end{figure*}

Figure 2 shows the AGN/galaxy cross-correlation functions compared to the galaxy/galaxy auto-correlation function. The left hand panel of Figure~\ref{fig:AGN_all_all_400000_4as_log.eps} shows that the X-ray AGN reside in more dense environments than the general galaxy population over all scales within the redshift range 1.0 $\leq z\,\leq$ 1.5. The errors on w($\theta)$ were calculated using bootstrap techniques using the same method as \cite{2006MNRAS.368L..20G}. A random sample of the X-ray AGN were chosen (with replacement) to make up a bootstrap catalogue which is the same size as the actual X-ray AGN catalogue. The cross-correlation of the bootstrap catalogue was then found, and the whole process was repeated 100 times. The standard deviation from 100 cross-correlation results then represents the 1$\sigma$ error on w$(\theta)$. The bootstrap errors in this case are slightly larger than the those derived from Poisson statistics.

This calculation was repeated with different samples of X-ray AGN to test for any biases in the data and to check for consistancy in the results. This was also done to test whether the flux limit imposed (see Section 2.2) had an effect on the magnitude of the cross-correlation, as the central pointing of the X-ray data is deeper and can therefore detect fainter AGN which may reside in different environments. Firstly, only those AGN with spectroscopic redshifts (of which 37 out of 62 of the X-ray sample have this data available) were cross-correlated with all galaxies in the UDS. There was no deviance from the results shown in the left hand panel of Figure~\ref{fig:AGN_all_all_400000_4as_log.eps}. The only difference was the magnitude of the errors. Secondly, all of the AGN within the 100ks pointing of the SXDS (before any flux cuts were made) were cross-correlated with all UDS galaxies within the same reduced area in order to make sure that there is no bias towards the AGN which are brighter in the X-ray waveband. Again, there were no changes to the main result of this paper other than the size of the errors involved. The final check was to cross-correlate the AGN within the 100ks pointing with all of the galaxies in the UDS; not only those contained within the same reduced area. No difference other than the magnitude of the errors was detected.

All three of these checks were repeated with every cross-correlation performed, including the cross-correlations of hard and soft X-ray AGN, optical and X-ray dominated AGN and finally the cross-correlations of X-ray AGN with the passive and star-forming populations (which are defined in Section 3.3). No deviance from any of the results displayed in this paper were found, and therefore the only results shown in this paper are those with the full sample of galaxies; i.e. all 62 AGN and all of the galaxies between 1.0 $\leq z\,\leq$ 1.5 in the UDS other than the areas used to mask stars and boundaries.

A closer look at the number of galaxies within close proximity to an X-ray AGN reveals that out of 62, 11 (17.7$\%$) have a nearby galaxy (within 5"; 42kpc) in the redshift range 1.0 $\leq z\,\leq$ 1.5 and with $K<23$. The same calculation was repeated for the general galaxy population of which 16.6$\%$ have a close galaxy partner. These results do not indicate an excess of very close companions around X-ray AGN, which may suggest that close interactions are not responsible for AGN triggering. However, this does not rule out merging as a mechanism, as we cannot resolve the internal structure of many of the AGN.

\subsubsection{Hard-Soft X-ray AGN}

The X-ray detected AGN were divided into two categories (hard and soft) by calculating the hardness ratio. We used the equation

\begin{equation}
HR = \frac{h - s}{h + s},
\end{equation}

where {\it h} is the flux in the hard band ($2-10\,{\rm keV}$) and {\it s} is the flux in the soft band ($0.5-2\,{\rm keV}$). A HR value of less than -0.4 is classified as soft as this corresponds to an AGN at $z=1$ with a moderate intrinsic absorbing column of $N_{H}=1\times10^{22}\,{\rm{~atoms~cm^{-2}}}$. This assumes an unobscured power law spectrum with photon index $\Gamma=2$. This value of HR also matches the natural division between soft sources at bright fluxes and the harder AGN emerging at faint levels. In our sample, 30 AGN were classified as hard, and 32 AGN as soft. No significant difference was found in the cross-correlation between the two categories of AGN, which can be seen in the left hand panel of Figure~\ref{fig:hard_soft_optical_Xray.eps}, which is in agreement with results from \cite{2009ApJ...701.1484C} at a similar redshift. This is in support of the AGN unification model which proposes that the only difference between the two AGN types is the orientation of the AGN with respect to the observer rather than any physical process. 

Following the same procedures that are detailed in section 4, the small scale environments of both types of AGN were investigated. The colour and the K-band magnitude of the neighbours of both AGN type were compared on 100kpc, 200kpc and 500kpc scales and K--S tests were performed to see if the neighbours were drawn from different parent populations. All tests confirmed that the neighbours of both types of AGN are consistent with being drawn from the same population. Again, this is consistent with the theory of AGN unification.

\subsubsection{Quasar-like and Seyfert-like AGN}

The X-ray AGN were divided into those in which infrared light from the galaxy itself dominates, and those which are dominated by X-ray flux from the AGN in order to compare those AGN which are quasar-like and those which are Seyfert-like. This involved calculating a value of $\nu \mathrm{f} \nu$ in both the K-band and in the (hard) X-rays for each AGN, as these wavebands are not as susceptible to gas and dust obscuration. They should therefore give a good representation of the AGN spectral energy distribution. These two values were then used to calculate a ratio between the the K-band flux and the X-ray flux. Those AGN that were defined to be optically-dominated (Seyfert-like AGN) had a K-band to X-ray flux ratio of greater than 2 (which is where there is a distinct turn-over in a plot of the flux ratios); and those with a ratio under 2 were defined to be X-ray dominated, which represents those galaxies which are more quasar-like. The mean K-band luminosity of the Seyfert-like galaxies was found to be $M_{k} = -26.2$. Using this method, 24 AGN were defined to be dominated by infrared light and the remaining 38 dominated by X-ray light. 

\begin{figure*}
\begin{center}
\hspace*{-20pt}
\includegraphics[width=15.0cm]{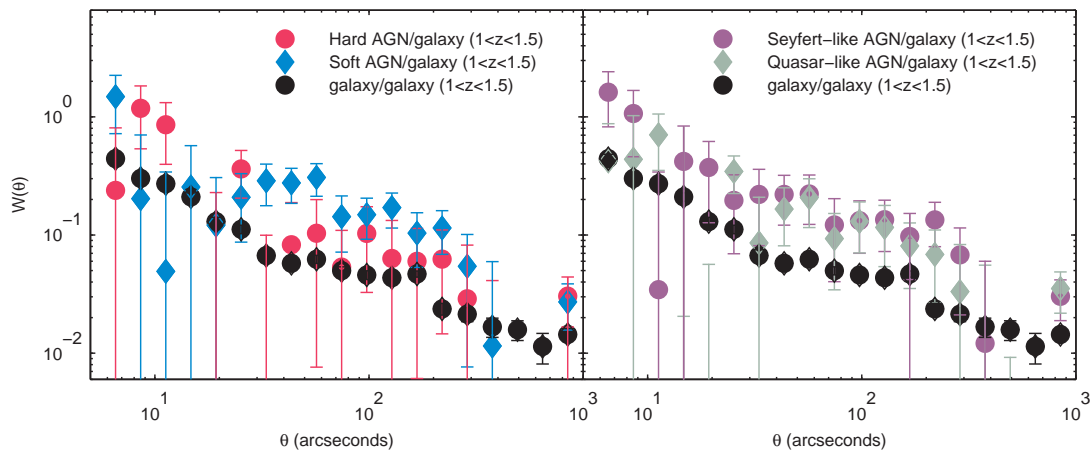} 
\end{center}
\caption{\small{The angular correlation functions for X-ray selected AGN between a redshift of 1.0 $\leq z\,\leq$ 1.5 with all galaxies within the same redshift. Left: The AGN have been split into two categories, hard and soft. Right: The X-ray AGN have been divided into those which are dominated by K-band flux (Seyfert-like AGN) and those that are dominated by X-ray flux (quasar-like AGN).}}
\label{fig:hard_soft_optical_Xray.eps}
\end{figure*}

No difference was found between the two populations (see Figure~\ref{fig:hard_soft_optical_Xray.eps}, right-hand panel). However, larger samples of AGN would be needed in order to investigate this result further, as well as the environments of hard and soft AGN (section 3.1.1) as small number statistics starts to play an important part in whether the cross-correlations are reliable.

\subsection{Radio-loud AGN correlations}\label{Radio_AGN_correlation}

There are 72 AGN in the UDS between a redshift of 1.0 $\leq z\,\leq$ 1.5 that have been detected in the radio using the data from \cite{2006MNRAS.372..741S}. Of these AGN, 18 have a nearby galaxy as defined in Section~\ref{Xray_AGN_correlations}. This represents 25 per cent of radio-loud AGN that could be associated with merging. This is not significantly higher than that of the general galaxy population which would again suggest that close interactions of galaxies are not the dominant mechanism in the activation of radio-loud AGN.

The right hand panel of Figure~\ref{fig:AGN_all_all_400000_4as_log.eps} shows the radio-loud AGN cross-correlated with all galaxies in the redshift range 1.0 $\leq z\,\leq$ 1.5, along with the galaxy-galaxy autocorrelation function at the same redshift. This shows that radio-loud AGN are living in dense environments, which is similar to what we see at lower redshifts \citep{2009ApJ...696..891H}.

In order to determine whether there was a dependence on radio luminosity with environment, the radio catalogue was divided into two (roughly equal) samples; those with a radio flux above 200$\mu$Jy and those below, with 35 and 37 AGN in each sample respectively. There was found to be no difference in the cross-correlation of either sample with the general galaxy population at 1.0 $\leq z\,\leq$ 1.5.

\subsection{Passive and Star-forming Galaxies}

\begin{figure*}
\begin{center}
\includegraphics[width=15.0cm]{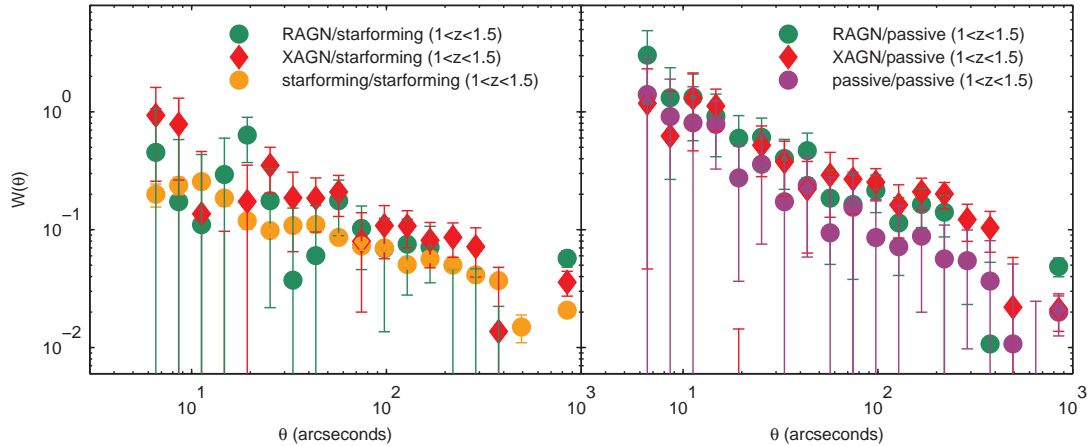} 
\end{center}
\caption{\small{The cross-correlation of X-ray AGN and radio-loud AGN with starburst (left) and passive (right) galaxies within the redshift range 1.0 $\leq z\,\leq$ 1.5.}}
\label{fig:passive_starburst}
\end{figure*}

The left panel of Figure~\ref{fig:passive_starburst} shows the angular cross-correlation of star-forming galaxies with both X-ray and radio-loud AGN, whilst the right hand panel of Figure~\ref{fig:passive_starburst} shows the same cross-correlation but with passive galaxies rather than star-forming galaxies. The autocorrelation functions for passive galaxies and star-forming galaxies are also shown on the figure for comparative purposes. The passive sample in this case was defined as those galaxies that are red in $U-B$ colour, to have a stellar age older than 1Gyr and have an ongoing star-formation rate of $<$0.5$\%$ of the initial rate implied by template fitting. The star-forming sample have only one criterion, and that is that they have a star-formation rate $>$10$\%$ than that of their initial star-formation rate. For a detailed definition of how the passive and starforming samples were formed see \cite{2010MNRAS.tmp.1089H}. Both types of AGN seem to be more strongly clustered with those galaxies that are passive rather than those galaxies which are actively forming stars.

This is perhaps not a surprising result. As shown in \cite{2008MNRAS.391.1301H}, passive populations of galaxies cluster more than star-forming galaxies and are among the most strongly clustered populations in the Universe at these redshifts. Therefore if an AGN is associated with a dense environment it is more likely to be associated with passive galaxies than star-forming galaxies.

\subsection{Halo Masses and Correlation Lengths}

Assuming that galaxies and AGN both trace the same underlying dark-matter distribution, in principle one can combine the AGN/galaxy cross-correlation function ($w_{AG}$) with the galaxy/galaxy auto-correlation function ($w_{GG}$) to infer the AGN/AGN auto-correlation function ($w_{AA}$).  Assuming  both populations trace the dark matter with linear bias, one can show that

\begin{equation}
w_{AA}=w_{AG}^2/w_{GG}.
\end{equation}

\begin{figure*}
\begin{center}
\includegraphics[width=15.0cm]{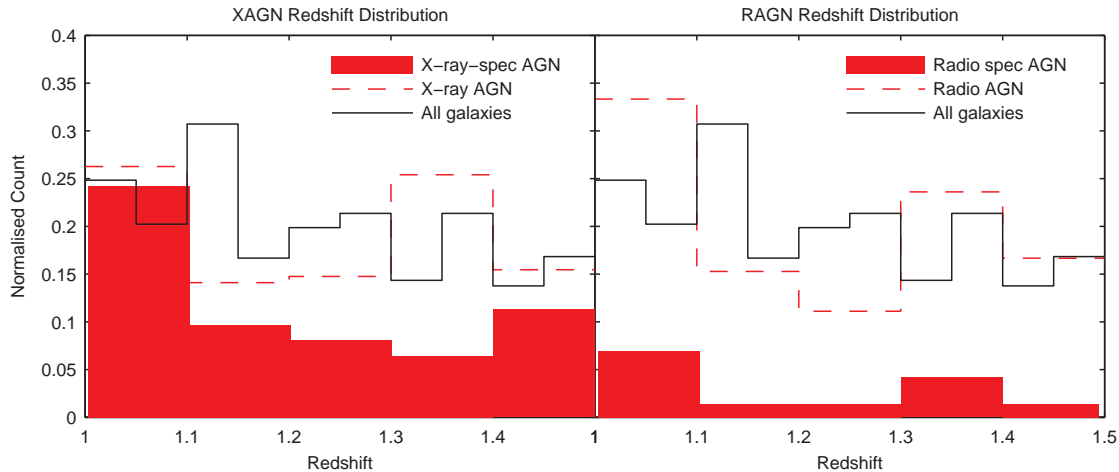} 
\end{center}
\caption{\small{The redshift distributions between 1.0 $\leq z\,\leq$ 1.5 of the total galaxy population (black), the overall AGN distribution (red, dashed), and those AGN with spectroscopic redshifts (filled red). Both the overall galaxy distribution and the AGN distribution have been normalised so that the area under each histogram is equal to 1. The filled red histogram has been normalised so that it represents the fraction of AGN with spectroscopic redshifts. The X-ray AGN are shown on the left panel, and the radio AGN are shown on the right panel.}}
\label{fig:redshift_distribution}
\end{figure*}

Using galaxies as tracers in this way can help to overcome the small number of AGN pairs to derive a more reliable estimate of their intrinsic clustering.  With projected angular correlation functions, however, we must also assume that both populations trace the same redshift distribution, which is broadly the case in our data over the redshift range $1.0<z<1.5$. This can be seen in Figure \ref{fig:redshift_distribution}. We used a two-sample Kolmogorov-Smirnov test (KS test) on the redshift distributions of both the spectroscopic and overall AGN populations with the null hypothesis that each of the two samples are drawn from the same continuous distribution as the general galaxy population. This was repeated for both the X-ray detected AGN and the radio AGN. The results show that all populations are consistent with being drawn from the same continuous distribution, and in no cases could the null hypothesis be rejected at a significance of $>95$ \%. We therefore find no strong evidence that the AGN and galaxy populations do not trace the same redshift distribution.

Figure 2 shows that the cross-correlation of radio and X-ray AGN with the full galaxy sample is significantly stronger than the galaxy/galaxy auto-correlation. From the equation above, the implied amplitude of AGN clustering must therefore be very high. This can also be seen in the right-hand panel of Figure 4, where the cross-correlation with passive galaxies is shown to yield an amplitude comparable to the passive galaxy auto-correlation function (and is in fact somewhat higher).

From the inferred magnitude of the XAGN/XAGN auto-correlation function, $w_{AA}$, we can fit a single power law of the form $w_{AA} = A(\theta^{-\delta} - C)$ to the data over the separation range $0.025 - 0.30$ degrees, which corresponds to a lower limit of 750kpc. We fix the slope at $\delta=0.8$ and minimise the $\chi^2$. We do not use the data below 0.025 degrees because it is known that the 1-halo term begins to dominate at this scale, as is evident from the cross-correlation statistics (see Figure \ref{fig:AGN_all_all_400000_4as_log.eps}). This deviation is due to multiple occupation of dark matter halos and can be modelled by a halo occupation distribution \citep[e.g.][]{{2002PhR...372....1C},{2004MNRAS.355.1010P},{2009MNRAS.397.1862P},{2011ApJ...726...83M}}. Such an analysis is, however, beyond the scope of this work. We do not fit to scales above 0.3 degrees as the size of the survey limits our measurements. The real space clustering and projected clustering are linked by the relativistic Limber equation \citep{1954ApJ...119..655L}.  If the redshift distribution of a sample is known, the Limber equation can be inverted and the
correlation length, $r_0$, can be calculated in a robust manner \citep{{1980lssu.book.....P},{1999MNRAS.306..988M}}. Taking the redshift distribution of the X-ray detected AGN, we obtain an estimate of the correlation length $r_0 = 17.0^{+2.3}_{-2.6}$ Mpc. The quoted errors are due to the error in the fit and therefore take into account bootstrap errors, but not those due to cosmic variance. We then use the method and formulae used by \cite{2002MNRAS.336..112M} \citep[which includes work from][]{{1974ApJ...187..425P},{1991ApJ...379..440B},{1996MNRAS.282..347M},{1998ApJ...503L...9J},{2001MNRAS.323....1S}} to calculate the typical halo mass within which the X-ray AGN reside. We compare the value of $r_0$ from our measurements to those of dark matter halos, which corresponds to a mass of $5.7^{+2.4}_{-2.2}\times 10^{13}$~M$_{\sun}$ at $z\, \sim$ 1.25.

We then repeated the same calculations on the inferred auto-correlation function of the radio AGN. We find a correlation length of $r_0 = 7.4^{+2.3}_{-3.0}$ Mpc, which corresponds to a typical halo mass of $3.2^{+6.0}_{-3.0}\times 10^{12}$~M$_{\sun}$ at $z\, \sim$ 1.25. This is indicative of both X-ray and radio selected AGN residing in significantly overdense environments compared to ordinary galaxies at the same epoch.

\subsection{Massive Galaxies}

An interpretation of our results so far could be that the AGN-galaxy cross-correlation is stronger than the galaxy autocorrelation because the UDS is very deep and therefore could be detecting many low-luminosity galaxies that are not as strongly clustered as those hosting AGN. In order to test whether this is the case, we re-calculated the autocorrelation function of all galaxies whilst slowly increasing the K-band luminosity threshold of galaxies included in the sample. Starting with all galaxies with a K-band luminosity brighter than $M_{k} = -23$, and decreasing in steps of 0.5 mag, the autocorrelation function was re-calculated until only those galaxies with a luminosity brighter than $M_{k} = -25$ remained. The magnitude of the autocorrelation function does not change until a K-band luminosity of $M_{k} = -25$ is reached, which mostly represents the most massive and passive galaxies. From calculating these autocorrelations, we can say that the low-luminosity galaxies in the UDS are not artificially lowering the amplitude of the galaxy/galaxy autocorrelation function. These results are consistent with the clustering analysis presented in \cite{2010MNRAS.tmp.1089H}, which showed that the strength of galaxy clustering does not depend strongly on luminosity until the samples are dominated by the red, passive galaxies at the highest luminosities.

An obvious interpretation of the strong cross-correlation signals is that AGN at these redshifts reside in massive galaxies which in turn are found in the most dense environments within the most massive dark-matter halos. This would, however, imply that X-ray AGN at $z>1$ reside in very different environments to those observed in the local Universe. It is important to note that although the host galaxies may be more massive, the luminosities of the AGN are still typical of those associated with the knee of the luminosity function \citep[][]{{2009A&A...500..749E},{2000A&A...353...25M}}, see Section 2.2. As was mentioned in Section~\ref{Introduction}, X-ray AGN in the local Universe reside in many different types of environment, and it appears that AGN live in star-forming galaxies irrespective of environment at $z<1$ \citep{2009ApJ...695..171S}.

\subsection{A Low-Redshift Comparison ($0.5 \leq z \leq 1.0$)}

Cross-correlations were also calculated between X-ray AGN, radio AGN and all galaxies for a lower redshift of 0.5 $\leq z\,\leq$ 1.0 in order to confirm findings from previous studies. Earlier work has found that X-ray AGN reside in moderately dense environments, as previously seen by \citep[][]{{2007MNRAS.380.1467G}, {2009ApJ...696..891H}, {2009ApJ...695..171S}}, whereas radio AGN prefer to reside in much denser regions associated with cluster environments \citep[][]{{2009ApJ...696..891H}, {2008MNRAS.391.1674W}}.

\begin{figure*}
\begin{center}
\hspace*{-20pt}
\includegraphics[width=15.0cm]{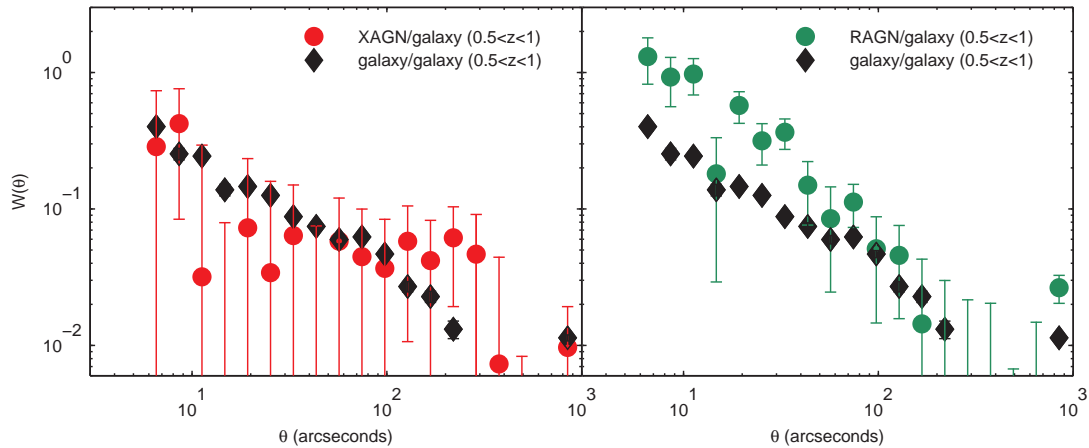} 
\end{center}
\caption{\small{The angular cross-correlation function for X-ray selected AGN between a redshift of 0.5 $\leq z\,\leq$ 1.0 compared to the galaxy-galaxy auto-correlation function within the same redshift.}}
\label{fig:AGN_radio_all_400000_4as_0_5_1_log.eps}
\end{figure*}

There were 60 X-ray AGN in the sample once the same selection techniques outlined in section ~\ref{Xray_AGN_correlations} were applied, along with 12,602 field galaxies (after masking and in the redshift range 0.5 $\leq z\,\leq$ 1.0). Of the X-ray AGN, 28 have spectroscopic redshifts and the remaining 32 have photometric redshifts. For a brief summary of the different samples and numbers of AGN in the low-redshift comparison, see Table 2. The mean X-ray luminosity in this sample of AGN is $7 \times 10^{43}\, {\rm erg~s^{-1}}$, which falls within the range of luminosities of X-ray selected AGN that were discussed by \cite{2009ApJ...696..891H} at a similar redshift. We are therefore probing X-ray AGN of similar luminosities to the 1.0 $\leq z\,\leq$ 1.5 sample, again around the knee in the luminosity function \citep[][]{{2009A&A...500..749E}, {2000A&A...353...25M}}. The result of the cross-correlation can be seen in the left panel of Figure~\ref{fig:AGN_radio_all_400000_4as_0_5_1_log.eps}. It is difficult to interpret the result, as the errors associated with the calculation are large due to the small number of AGN, but the similarity to the galaxy/galaxy autocorrelation function suggests that the AGN reside in environments comparable to ordinary galaxies at these redshifts. However, the distinction is not as strong as that seen in Figure~\ref{fig:AGN_all_all_400000_4as_log.eps}, suggesting that there is some kind of evolution associated with X-ray AGN.

\begin{table}
\centering
\begin{tabular}{l c c}
\hline\hline
Sample & Number & No. with spec-z \\ [0.5ex]
\hline
X-ray AGN & 60 & 28 \\
Radio AGN & 64 & 12 \\
\hline
Galaxy Sample & 12\,602 & - \\
\hline\hline
\end{tabular}
\label{table:table2}
\caption{A table of the AGN and galaxy samples in the redshift range 0.5 $\leq z\,\leq$ 1.0, along with the number of galaxies in each sample and the number that have spectroscopic redshifts.}
\end{table}

Unlike the previous radio AGN sample in which we assumed all sources were AGN rather than star-forming galaxies, the radio sources at the redshift of 0.5 $\leq z\,\leq$ 1.0 were divided into two categories; those which are star-forming and those which are AGN. The AGN luminosity threshold was determined by setting a maximum star-formation rate, converting this to a nonthermal luminosity $L_{N}$ and consequently a 1.4GHz flux (assuming an average redshift of $z=0.75$) using the following equation from \cite{1992ARA&A..30..575C}:

\begin{equation}
\left(\frac{L_{N}}{\mathrm{WHz}^{-1}}\right) \sim 5.3\times10^{21}\left(\frac{\nu}{\mathrm{GHz}}\right)^{-\alpha}\left[\frac{SFR(M\geq5M_{\sun})}{M_{\sun}\mathrm{yr}^{-1}}\right],
\end{equation}

where $\alpha$ is the nonthermal spectral index. There is also a radio thermal fraction to consider, but at a frequency of 1.4GHz, most of the radio signal will be from synchrotron emission, and therefore it is the nonthermal contribution which will dominate.

There will be some overlap in radio luminosity between star-forming galaxies and AGN; however, by setting a high maximum star-formation rate (we set a threshold at 100 solar masses per year), this can be kept to a minimum. In this way we obtain a sample of 64 radio AGN in the redshift range 0.5 $\leq z\,\leq$ 1.0, of which 12 have spectroscopic redshifts. The right hand panel of Figure~\ref{fig:AGN_radio_all_400000_4as_0_5_1_log.eps} shows that at this epoch radio AGN do indeed reside in very dense environments. This is consistent with radio AGN being found in the dense regions associated with the centre of galaxy clusters, as found by \cite{2002MNRAS.330...17B} and \cite{2003MNRAS.343..924J} at a similar redshift.

Overall, our results appear to confirm previous work at low redshift; we find that radio AGN at $z<1$ reside in significantly denser environments than X-ray emitting AGN.

\section{Small Scale AGN Environments}\label{science}

\begin{figure*}
\begin{center}
\includegraphics[width=18.0cm]{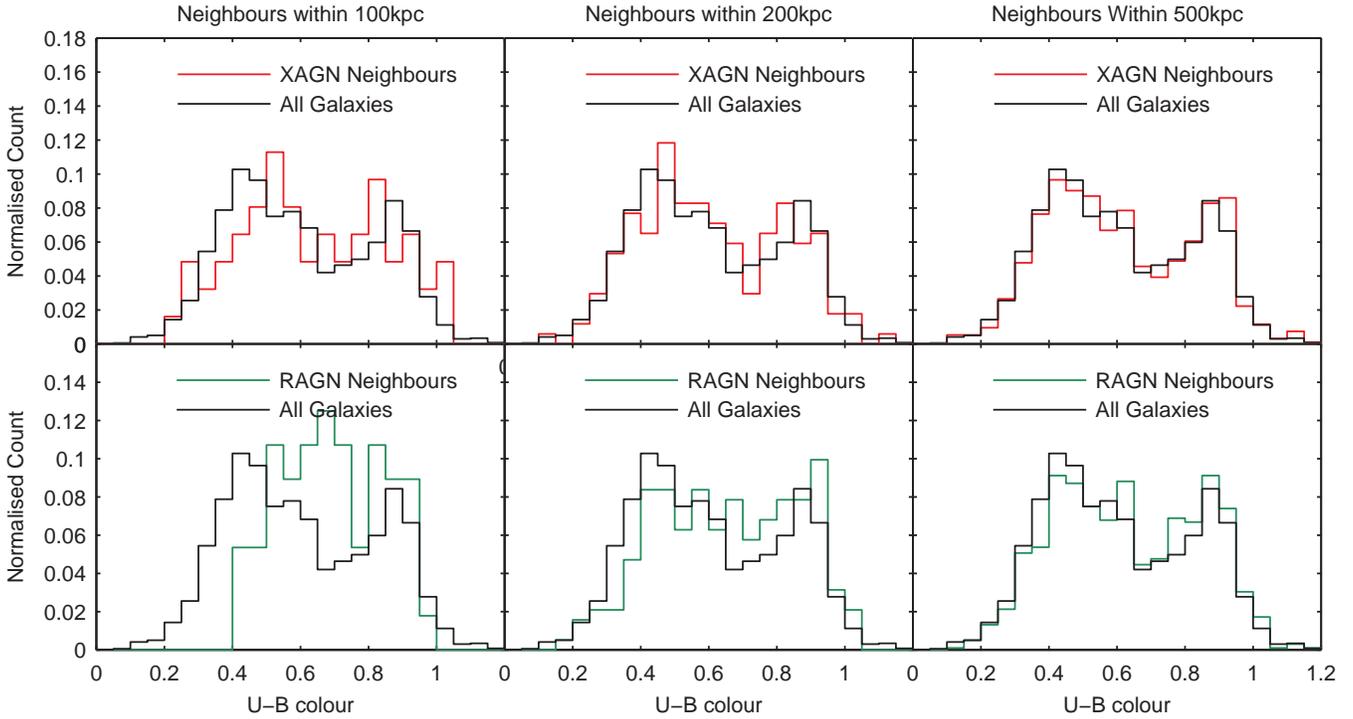} 
\end{center}
\caption{\small{The $U-B$ colour of the neighbours within 100kpc, 200kpc and 500kpc of X-ray and radio AGN within the redshift 1.0 $\leq z\,\leq$ 1.5 compared to the general population.}}
\label{fig:Colour_of_neighbours}
\end{figure*}

The cross-correlations at high redshift (1.0 $\leq z\,\leq$ 1.5) in Section 3 show that both types of AGN are found in dense environments. In order to investigate whether there are differences in the type of environment, we studied the properties of the immediate galaxies surrounding the AGN. We made three separate samples of the neighbouring galaxies. The first sample includes those that are within a physical distance of 100kpc from an AGN and should therefore include very close neighbouring galaxies. A second sample was drawn of those galaxies that are found within 200kpc of an AGN which should represent its group environment. The final category was made from those galaxies that are within 500kpc of an AGN, which should be representative of the wider environment. These three categories were then each divided into two further samples representing the neighbouring galaxies to an X-ray AGN and to a radio-loud AGN.

We then compared the properties of the galaxies contained within these six samples with the overall galaxy population, starting with $U-B$ colour (see Figure 6). We used a KS test of the $U-B$ colour distribution with the null hypothesis that each of the six samples are drawn from the same continuous distribution as the general galaxy population. Comparing 100kpc, 200kpc and 500kpc samples of galaxies surrounding the X-ray AGN with the general galaxy population shows that the two populations are consistent with being drawn from the same continuous distribution, rejecting the null hypothesis at only 50.8$\%$, 10.9$\%$ and 42.1$\%$ significance respectively. A KS test of the radio-loud AGN neighbours at 100kpc, 200kpc and 500kpc with the general population reveals however that the two populations are drawn from different distributions, rejecting the null hypothesis at 99.994$\%$, 99.98$\%$ and 99.8$\%$ significance.

KS tests were also run on the absolute K-band magnitude of the same galaxy samples. All populations were found to be drawn from the same data set, with the exception of the neighbours within a 200kpc radius of a radio-loud AGN. These galaxies were found to be drawn from a different (typically brighter) population of galaxies to 98.4$\%$ significance.

These results show that radio-loud AGN live in different environments to the general galaxy population. From Figure 6, it seems as though radio AGN are associated with galaxies that are red or green in $U-B$ colour. Combining these results with those from Section 3 where radio-loud galaxies are found to be in dense environments at a redshift 1.0 $\leq$z$\leq$ 1.5 shows that radio galaxies probably live in very similar environments to which they are found in the local Universe.

Inspecting the optical/IR images by eye, it is indeed possible to see that the radio AGN are associated with overdense areas, although such conclusions are somewhat qualitative. In a large fraction of cases, the radio AGN can be seen to be in the centre of small clusters of galaxies, and seem to be especially associated with those galaxies that are red in $U-B$ colour. This suggests that radio-loud AGN at this redshift can still tentatively be associated with dense groups of galaxies, similar to what is seen in the near Universe. A similar inspection of the X-ray selected AGN reveals that the range of environments they reside in seems to vary; a significant amount (again, around 50$\%$) seem to be associated with the larger structures in this redshift range, although they are not so predominantly found in the centres of these structures.

\section{Discussion and Conclusions}\label{discussion}

We investigate the environments of both X-ray and radio-loud AGN within the UKIDSS Ultra-deep survey (UDS), focusing on the redshift range $1.0 \leq z\,\leq 1.5$. Deep K-band selection is used to sample the galaxy density field, using $9896$ galaxies with photometric
redshifts derived from 11-band optical/IR photometry. Our measure of `environment' is defined by projected galaxy overdensity, as measured by angular cross-correlation. We find that both X-ray and radio-loud AGN reside in significantly overdense environments compared to ordinary galaxies (Figure 2), with cross-correlation amplitudes suggesting environments comparable to luminous passive galaxies at this epoch.  Combining the cross-correlation results with the galaxy auto-correlation function we infer dark matter halo masses for X-ray emitting AGN of order $\sim 10^{13}\mathrm{M_{\sun}}$, comparable to the halos occupied by the progenitors of massive elliptical galaxies \citep{2010MNRAS.tmp.1089H}. It would be of benefit to obtain AGN and galaxy samples with a larger fraction of spectroscopic redshifts in order that more confidence can be placed in the cross-correlation analysis and so that errors on calculations made on these measurements can be reduced.

Our results suggest a change in the environments of X-ray emitting AGN at $z>1$ compared to lower redshifts.  Previous studies revealed that X-ray emitting AGN at $z<1$ live in environments comparable to blue starforming galaxies \citep[e.g.][]{2009ApJ...695..171S}, with a tendency to avoid high-density regions such as the central regions of galaxy clusters \citep[e.g.][]{{2007MNRAS.380.1467G}, {2009ApJ...700..901K}}. If confirmed, the very different environments we observe at $z>1$ may indicate a `downsizing' scenario in which AGN activity occurs preferentially within the most massive dark matter halos at high redshift, shifting to lower mass systems towards the present day. Such an evolutionary model was proposed by \cite{2009ApJ...696..891H}, who argue that an X-ray bright phase occurs when a galaxy's dark matter halo grows to a critical mass between $\sim 10^{12}\mathrm{M_{\sun}}$ and $\sim 10^{13}\mathrm{M_{\sun}}$, thereafter shifting to radiatively inefficient radio-mode activity.  Given our small sample (62 X-ray emitting AGN) and the relatively small size of the field ($\sim 40\times 40$ comoving Mpc at $z\simeq 1$) we believe that further studies will be required to test our findings, which may be influenced by cosmic variance.

Mergers and interactions between galaxies are also thought to play an important role in the type and frequency of AGN activity. Simulations have been able to reproduce the observed peak of high luminosity AGN at $z\, \sim$ 2 through major mergers of high mass galaxies
\citep{2008MNRAS.391..785W}. \cite{2007ApJ...660L..15G} find that X-ray AGN avoid underdense environments and therefore conclude that a SMBH becomes active during the infall of its host galaxy to an overdense region. They also suggest that lower luminosity AGN activity
could be triggered by interactions with neighbouring galaxies, but not due specifically to merging; a conclusion that \cite{2009ApJ...695..171S} also draw from their studies of AGN environments in the COSMOS field. Our results may indicate that such processes are more prevalent in higher mass halos at $z>1$ compared to the present day.

In contrast to what is found at lower redshifts, we find that X-ray AGN reside in significantly overdense environments at 1.0 $\leq z\,\leq$ 1.5. Although the errors involved are large, we also find that X-ray AGN are found in slightly more massive halos than those of radio AGN at this redshift. On small scales, however, we find that the neighbours of radio-loud AGN differ from those of X-ray AGN in rest-frame $U-B$ colour. On scales $<200$kpc the neighbours of radio-loud AGN are significantly skewed towards red sequence galaxies, while the neighbours of X-ray emitting AGN appear similar to the general galaxy population. This may suggest that radio-loud AGN are preferentially located in the cores of massive halos while X-ray emitting AGN are generally located in the outskirts, perhaps triggered by infall as suggested by \cite{2003MNRAS.343..924J} and \cite{2007ApJ...660L..15G}.  Alternatively, it may be that X-ray emitting AGN are typically located in more `active' halos, in which galaxies are undergoing more nuclear and star-forming activity generally, perhaps triggered by a recent halo merging event. Radio AGN may then be located in older systems, where star-formation has typically ceased and the more massive galaxies have entered radiatively inefficient, radio-loud modes of accretion, as suggested by \cite{2009ApJ...696..891H}. Further investigation with larger samples will be required to disentangle these possibilities.

\section*{Acknowledgments}

We are indebted to the staff at UKIRT for operating the telescope with
such care and dedication. We also thank the teams at CASU and WFAU for
processing and archiving the data.

\bsp

\label{lastpage}

\bibliographystyle{mn2e.bst}
\bibliography{bibliography3}

\end{document}